\documentclass[twocolumn,showpacs,preprintnumbers,floatfix,prl]{revtex4}
\usepackage{graphicx}

\begin{document}
\title{

Breaking of ergodicity and long relaxation times in systems with
long-range interactions}
\author{
D. Mukamel $^{1}$\email{fnmukaml@wisemail.weizmann.ac.il},
S. Ruffo $^{2}$\email{ruffo@avanzi.de.unifi.it}, and N. Schreiber $^1$}                     
\affiliation{$^1$ Department of Physics of Complex Systems,
Weizmann Institute of Science, Rehovot, Israel 76100 \\
$^2$ Dipartimento di Energetica ``Sergio Stecco", Universit\`a di
Firenze, INFN and CSDC, via s. Marta 3, 50139 Firenze, Italy. }



\date{\today}

\begin{abstract}

The thermodynamic and dynamical properties of an Ising model with
both short range and long range, mean field like, interactions are
studied within the microcanonical ensemble. It is found that the
relaxation time of thermodynamically {\it unstable} states
diverges logarithmically with system size. This is in contrast
with the case of short range interactions where this time is
finite. Moreover, at sufficiently low energies, gaps in the
magnetization interval may develop to which no microscopic
configuration corresponds. As a result, in {\it local
microcanonical dynamics} the system cannot move across the gap,
leading to breaking of ergodicity even in finite systems. These
are general features of systems with long range interactions and
are expected to be valid even when the interaction is slowly
decaying with distance.

\end{abstract}


\maketitle

Systems with long range interactions are rather common in
nature~\cite{LesHouches}. In such systems the inter-particle
potential decays at large distance $r$ as $1/r^\alpha$ with
$\alpha \le d$, in dimension $d$. Examples include magnets with
dipolar interactions, wave-particle interactions~\cite{Barre},
gravitational forces~\cite{Padmanabhan}, and Coulomb forces in
globally charged systems~\cite{Nicholson}. It is well known that
in such systems the various statistical mechanical ensembles need
not be equivalent~\cite{Thirring}. For example whereas the
canonical ensemble specific heat is always non-negative, it may
become negative within the microcanonical ensemble when long range
interactions are present~\cite{Lynden68}. It has recently been
suggested that whenever the canonical ensemble exhibits a first
order phase transition the canonical and the microcanonical phase
diagrams may be different~\cite{Mukamel01}. This has been
demonstrated by a detailed study of a spin-1 Ising model with long
range, mean field like, interactions. While the thermodynamic
behavior of such models is fairly well understood, their dynamics,
and the approach to equilibrium, has not been investigated in
detail so far~\cite{Yama}. The aim of the present Letter is to
identify some general dynamical features which distinguish systems
with long range interactions from those with short range ones.

One characteristic of systems with short range interactions is
that the domain in the space of extensive thermodynamic variables
$\vec{X}$ over which the model is defined is convex for
sufficiently large number of particles. Here the components of the
vector $\vec{X}$ are variables like the energy, volume and
possibly other extensive parameters corresponding to the system
under study. Convexity is a direct result of additivity. By
combining two appropriately weighted sub-systems with extensive
variables $\vec{X_1}$ and $\vec{X_2}$, any intermediate value of
$\vec{X}$ between $\vec{X_1}$ and $\vec{X_2}$ may be realized. On
the other hand systems with long range interactions are not
additive, and thus intermediate values of the extensive variables
are not necessarily accessible. This feature has a profound
consequence on the dynamics of systems with long range
interactions. Gaps may open up in the space of extensive variables
and lead to breaking of ergodicity under local microcanonical
dynamics of such systems. An example of such gaps in a class of
anisotropic $XY$ models has recently been discussed
in~\cite{Celardo}.

Another interesting feature of systems with long range
interactions is their relaxation time. It is well known that the
relaxation time of metastable states grows exponentially with the
system size \cite{Griffiths}. On the other hand the relaxation
processes of thermodynamically unstable states are not well
understood. In systems with short range interactions the
relaxation takes place on a finite time scale. However previous
studies of a mean field $XY$ model suggest that this relaxation
time diverges with the system size \cite{Yama}.

In the present Letter we study some of the issues discussed above
by considering a spin-$1/2$ Ising model with both long range, mean
field like, and short range nearest neighbor interactions on a
ring geometry~\cite{Nagle,Kardar}. We study its thermodynamic and
dynamical behavior in both the canonical and microcanonical
ensembles. It is found that the two ensembles result in different
phase diagrams as was observed in other models~\cite{Mukamel01}.
Moreover, we find that for sufficiently low energy, gaps open up
in the magnetization interval $-1 \le m \le 1$, to which no
microscopic configuration corresponds. Thus the phase space breaks
into disconnected regions. Within a {\it local} microcanonical
dynamics the system is trapped in one of these regions, leading to
a breakdown of ergodicity even in finite systems. In studying the
relaxation time of thermodynamically {\it unstable} states,
corresponding to local minima of the entropy, we find that unlike
the case of short range interactions where this time is finite,
here it diverges logarithmically with the system size. We provide
a simple explanation for this behavior by analyzing the dynamics
of the system in terms of a Langevin equation.

\begin{figure}[t]
\begin{center}
\includegraphics[width=6.00cm,height=4.00cm]{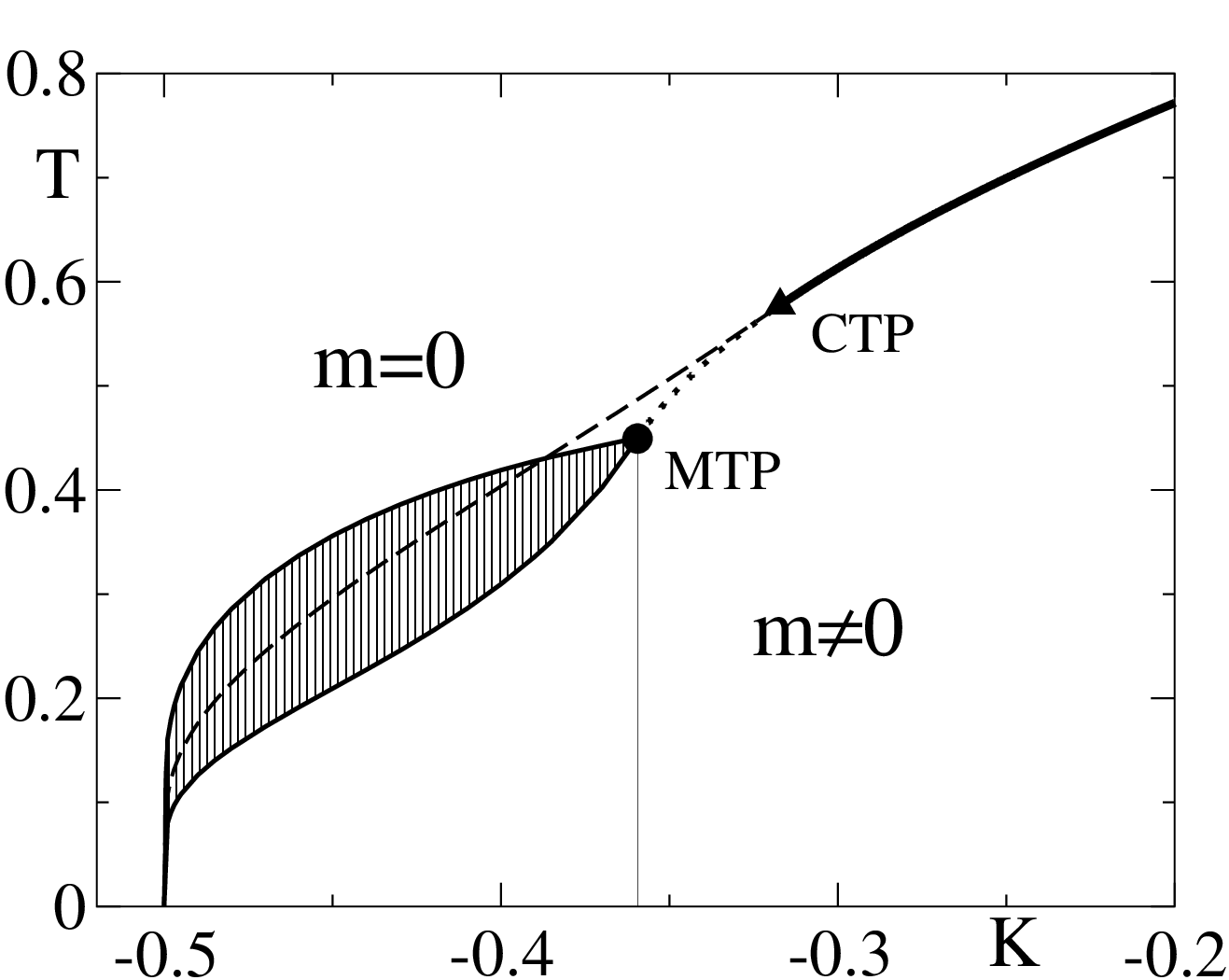}
\caption{\label{fig:phase_diagram} The canonical and
microcanonical $(K,T)$ phase diagram. In the canonical ensemble
the large $K$ transition is continuous (bold solid line) down to
the trictitical point CTP where it becomes first order (dashed
line). In the microcanonical ensemble the continuous transition
coincides with the canonical one at large $K$ (bold line). It
persists at lower K (dotted line) down to the tricritical point
MTP where it turns first order, with a branching of the transition
line (solid lines). The region between these two lines (shaded
area) is not accessible.}
\end{center}
\end{figure}

We start by considering an Ising model defined on a ring of $N$
spins $S_i=\pm 1$ with long and short range interactions. The
Hamiltonian is given by
\begin{equation}{\label{eq:Hamiltonian}}
H=-\frac{K}{2}{\sum_{i=1}^N}\left(S_iS_{i+1}-1\right)-\frac{J}{2N}
\left({\sum_{i=1}^N}S_i\right)^2,
\end{equation}
where $J>0$ is a ferromagnetic, long range, mean field like
coupling, and $K$ is a nearest neighbor coupling which may be of
either sign. The canonical phase diagram of this model has been
derived in the past~\cite{Kardar,Nagle}. It has been observed that
in the $(K,T)$ plane, where $T$ is the temperature, the model
exhibits a phase transition line separating a disordered phase
from a ferromagnetic one (see Fig.~\ref{fig:phase_diagram}). The
transition is continuous for large $K$, where it is given by
$\beta=e^{-\beta K}$. Here $\beta = 1/T$ and $J=1$ is assumed for
simplicity. Throughout this work we take $k_B=1$ for the Boltzmann
constant. The transition becomes first order below a tricritical
point located at an antiferromagnetic coupling $K_{CTP}=-\ln 3 /{2
\sqrt{3}}\simeq -0.317$.

We now analyze the model within the microcanonical ensemble. Let
$U=-\frac{1}{2}\sum_i\left(S_iS_{i+1}-1\right)$ be the number of
antiferromagnetic bonds in a given configuration characterized by
$N_+$ up spins and $N_-$ down spins. Simple counting yields that
the number of microscopic configurations corresponding to
$(N_+,N_-,U)$ is given, to leading order in $N$, by
\begin{equation}
\Omega(N_+,N_-,U) = \left( \begin{array}{c} N_+\\U/2
\end{array}\right )
\left( \begin{array}{c} N_-\\U/2
\end{array}\right )~.\\
\end{equation}
Expressing $N_+$ and $N_-$ in terms of $N$ and the magnetization
$M=N_+ - N_-$, and denoting $m=M/N$, $u=U/N$ and the energy per
spin $\epsilon=E/N$, one finds that the entropy per spin,
$s(\epsilon,m)=\frac{1}{N}\ln\Omega$ is given in the thermodynamic
limit by
\begin{eqnarray}{\label{eq:entropy}}
s(\epsilon,m)&=&\frac{1}{2}(1+m)\ln(1+m)+\frac{1}{2}(1-m)\ln(1-m)
\nonumber
\\&-& u\ln u -\frac{1}{2}(1+m-u)\ln(1+m-u)
\nonumber
\\& -&\frac{1}{2}(1-m-u)\ln(1-m-u)~,
\end{eqnarray}
where $u$ satisfies
\begin{equation}
\label{energy} \epsilon=-\frac{J}{2}m^2+Ku \,.
\end{equation}
By maximizing $s(\epsilon,m)$ with respect to $m$ one obtains both
the spontaneous magnetization 
$m_s(\epsilon)$ and the entropy $s(\epsilon)\equiv
s(\epsilon,m_s(\epsilon))$ of the system for a given energy. The
temperature, and thus the caloric curve, is given by $1/T=\partial
s(\epsilon) /\partial \epsilon$. A straightforward analysis of
(\ref{eq:entropy}) yields the microcanonical $(K,T)$ phase diagram
of the model (Fig.~\ref{fig:phase_diagram}), where it is also
compared with the canonical one. It is found that the model
exhibits a critical line given by the same expression as that of
the canonical ensemble. However this line extends beyond the
canonical tricritical point, reaching a microcanonical tricritical
point at $K_{MTP}\simeq -0.359$, which is computed analytically.
For $K<K_{MTP}$ the transition becomes first order and it is
characterized by a discontinuity in the temperature. The
transition is thus represented by two lines in the $(K,T)$ plane
corresponding to the two temperatures at the transition point.
These lines are obtained by numerically maximizing the entropy
(\ref{eq:entropy}).


We now consider the dynamics of the model. This is done by using
the microcanonical Monte Carlo dynamics suggested by
Creutz~\cite{Creutz}. In this dynamics one samples the microstates
of the system with energy $E-E_D$ with $E_D \ge 0$ by attempting
random single spin flips. One can show that, to leading order in the
system size, the distribution of the energy $E_D$ takes the form

%
%
\begin{equation}
\label{distribution}
P(E_D)\sim e^{-E_D/T}~.
\end{equation}
Thus, measuring this distribution, yields the temperature $T$
which corresponds to the energy $E$.

In applying this dynamics to our model one should note that to
next order in $N$ the energy distribution is given by $P(E_D)\sim
\exp{(-E_D/T-E_D^2/{2C_VT^2})}$, where $C_V=O(N)$ is the specific
heat. In extensive systems with short range interactions the
specific heat is positive and thus the correction term does not
modify the distribution function for large $N$. However in our
system $C_V$ can be negative in a certain range of $K$ and $E$. It
even becomes arbitrarily small near the MTP point, making the next
to leading term in the expansion arbitrarily large. This could in
principle qualitatively modify the energy distribution. However as
long as the entropy is an increasing function of $E$, namely for
positive temperatures, the distribution function
(\ref{distribution}) is valid in the thermodynamic limit. This
point is verified by our numerical studies, where an exponential
distribution of the energy $E_D$ is clearly observed.


We now address the issue of the accessible magnetization intervals
in this model. We find that in certain regions in the $(K,E)$
plane, the magnetization $m$ cannot assume any value in the
interval $(-1,1)$. There exist gaps in this interval to which no
microscopic configuration could be associated. To see this, we
take for simplicity the case $N_+>N_-$. It is evident that the
local energy $U$ satisfies $0\le U \le 2N_-$. The upper bound is
achieved in microscopic configurations where all down spins are
isolated. This implies that $0 \le u \le 1-m$. Combining this with
(\ref{energy}) one finds that for positive $m$ the accessible
states have to satisfy
\begin{eqnarray}
\label{Mrestriction}
&&m\leq \sqrt{-2\epsilon} \quad ,\quad  m \geq m_+ \quad ,\quad  m\leq m_- \nonumber \\
&&\mbox{with}\,\,  m_{\pm}=-K\pm \sqrt{K^2-2(\epsilon-K)}~.
\end{eqnarray}

\begin{figure}[ht]
\begin{center}
\includegraphics[width=6.00cm,height=4.00cm]{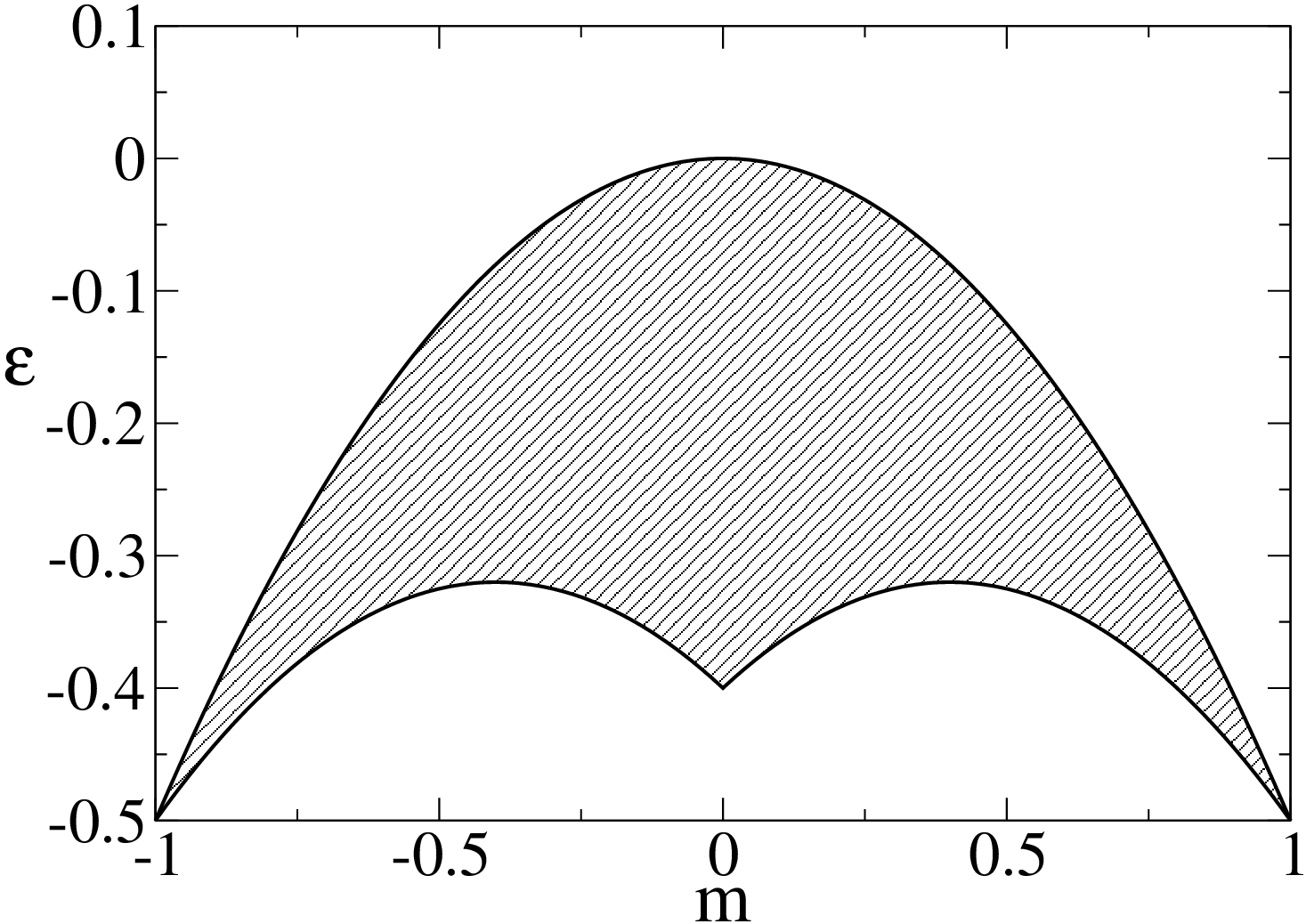}
\caption{\label{fig:Forbbiden_zone}Accessible region in the
$(m,\epsilon)$ plane (shaded area) for $K=-0.4$. For energies in a
certain range, gaps in the accessible magnetization values are
present.}
\end{center}
\end{figure}

\begin{figure}[t]

\begin{center}
\includegraphics[width=6.00cm,height=4.00cm]{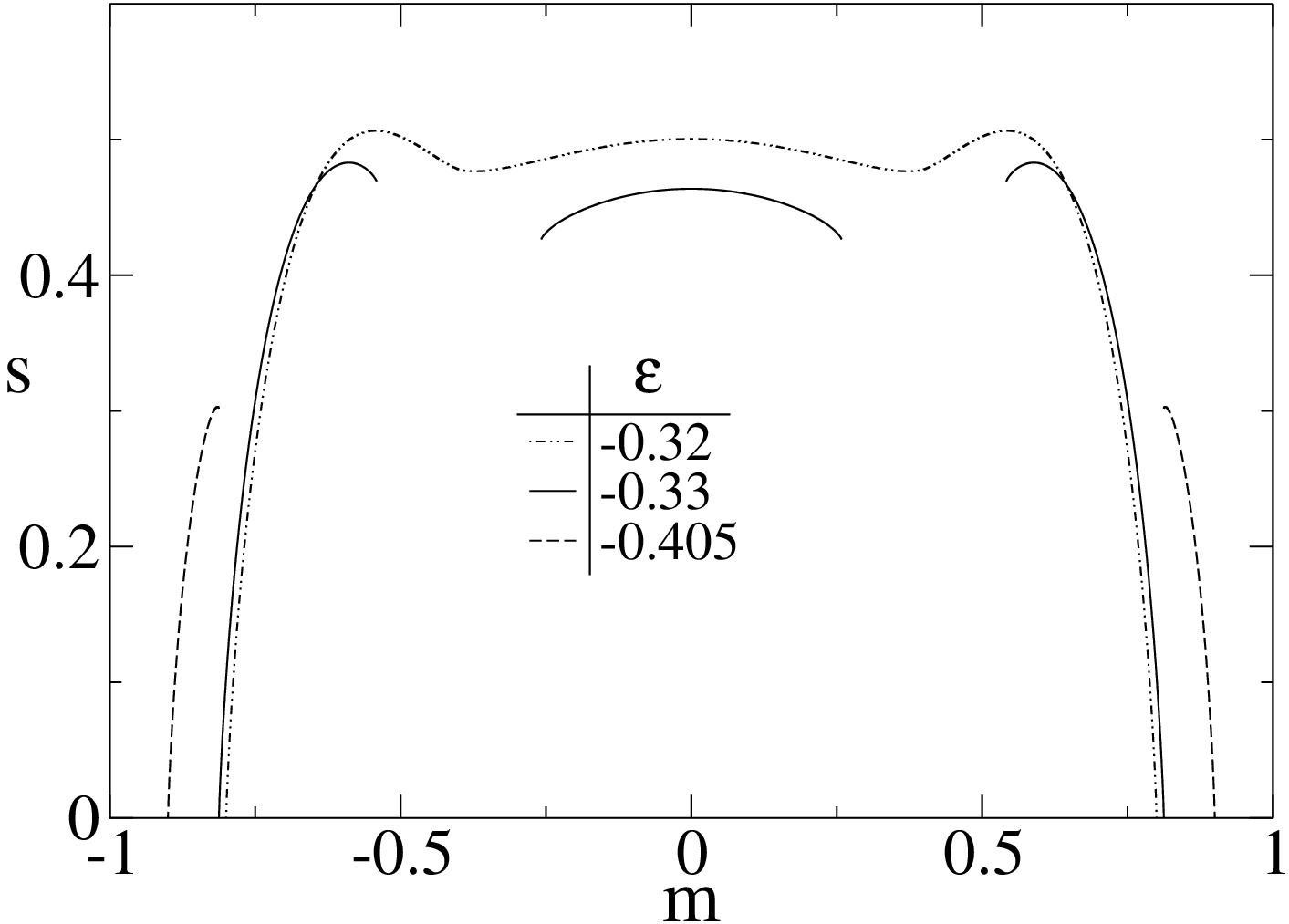}
\caption{\label{fig:s_vs_m} The $s(m)$ curves for $K=-0.4$, 
and for typical energy values, demonstrating that gaps in the
accessible states develop as the energy is lowered.}
\end{center}
\end{figure}

Similar restrictions exist for negative $m$. These restrictions
yield the accessible magnetization domain shown in Fig.
\ref{fig:Forbbiden_zone} for $K=-0.4$. It is clear that this
domain is not convex. Entropy curves $s(m)$ for some typical
energies are given in Fig. \ref{fig:s_vs_m}, demonstrating that
the number of accessible magnetization intervals changes from one
to three, and then to two as the energy is lowered.

This feature of disconnected accessible magnetization intervals,
which is typical to systems with long range interactions, has
profound implications on the dynamics. In particular, starting
from an initial condition which lies within one of these intervals,
local dynamics, such as the one applied in this work, is unable to
move the system to a different accessible interval. Thus
ergodicity is broken in the microcanonical dynamics even at finite
$N$.

\begin{figure}[ht]
\begin{center}
\includegraphics[width=7.00cm,height=5.50cm]{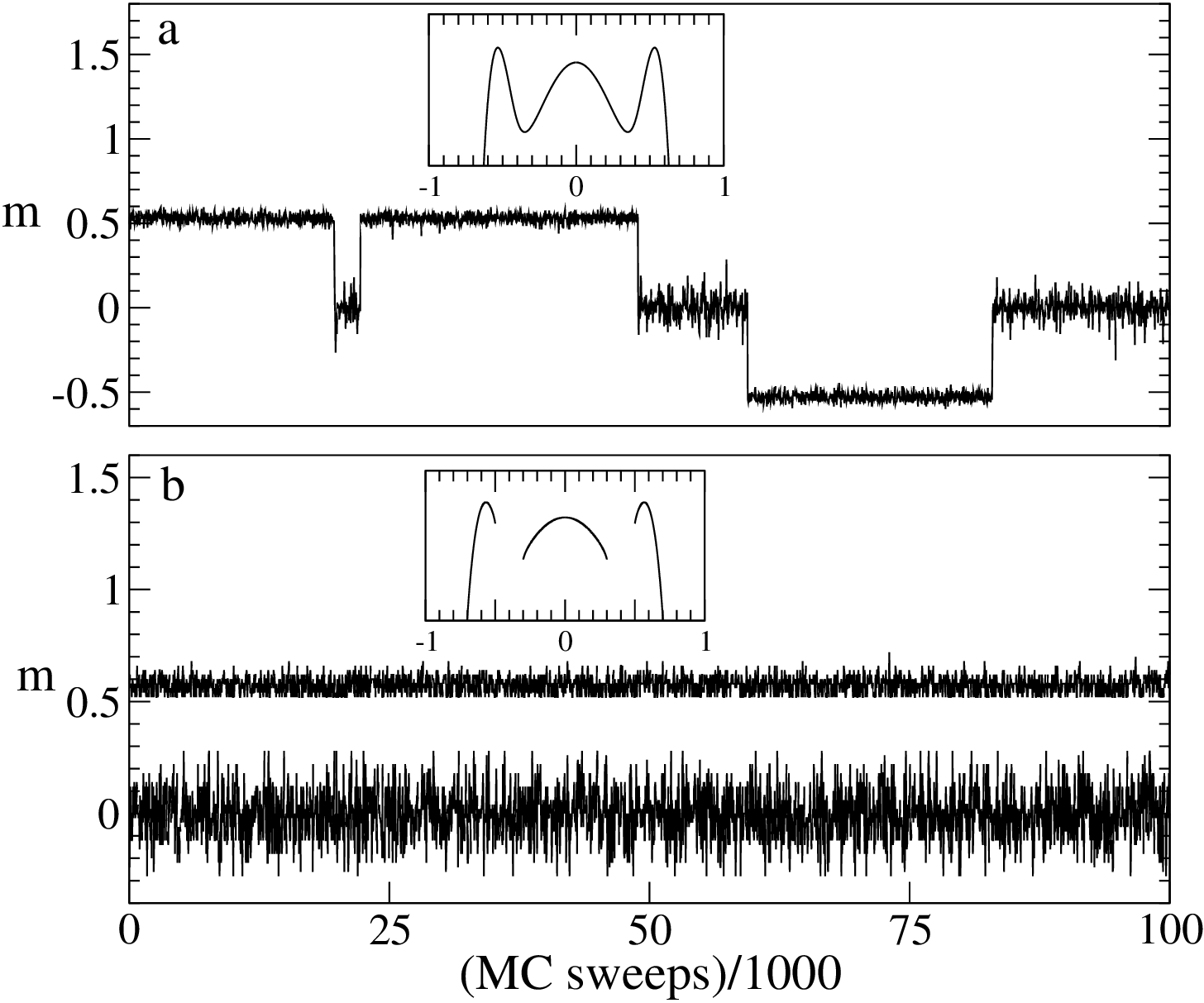}
\caption{\label{fig:m_vs_t}Time evolution of the magnetization for 
$K=-0.4$ (a) in the ergodic region ($\epsilon=-0.318$) 
and (b) in the non-ergodic region ($\epsilon=-0.325$). 
The corresponding entropy curves are shown in the inset.}
\end{center}
\end{figure}

To demonstrate this point we display in Fig. \ref{fig:m_vs_t} the
time evolution of the magnetization for two cases: one in which
there is a single accessible magnetization interval, where one
sees that the magnetization switches between the metastable $m=0$
state and the two stable states $m=\pm m_s$. In the other case the
metastable $m=0$ state is disconnected from the stable ones,
making the system unable to switch from one state to the other.
Note that this feature is characteristic of the microcanonical
dynamics. Using local canonical dynamics, say Metropolis algorithm
~\cite{Metropolis} , would allow the system to cross the forbidden
region (by moving to higher energy states where the forbidden
region diminishes), and ergodicity is restored in finite systems.
However, in the thermodynamic limit, ergodicity would be broken
even in the canonical ensemble, as the switching rate between the
accessible regions decreases exponentially with $N$.
\vspace{0.20cm}
\begin{figure} [h]
\begin{center}
\begin{tabular}{lr}
\includegraphics[height=3.0cm]{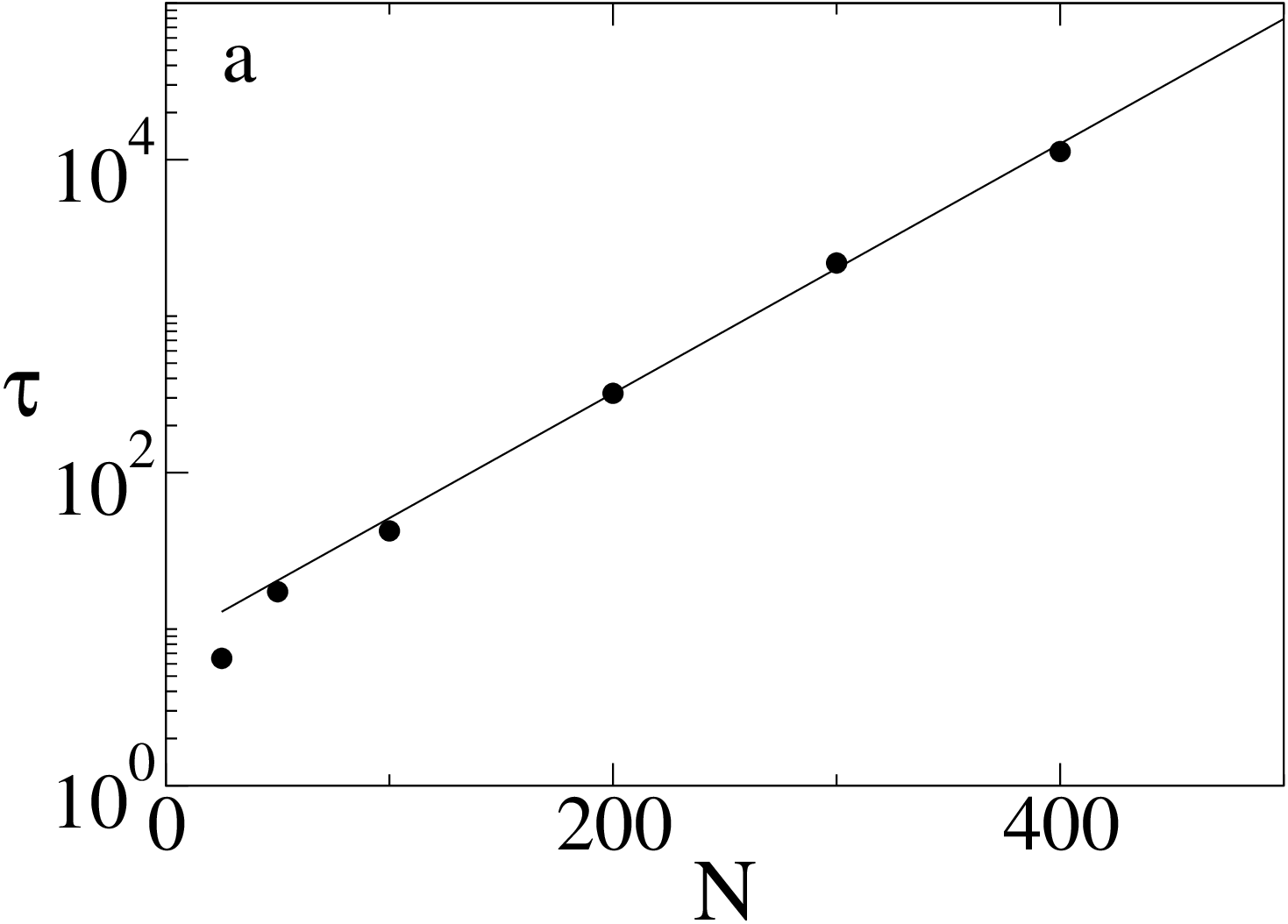}&
\includegraphics[height=3.0cm]{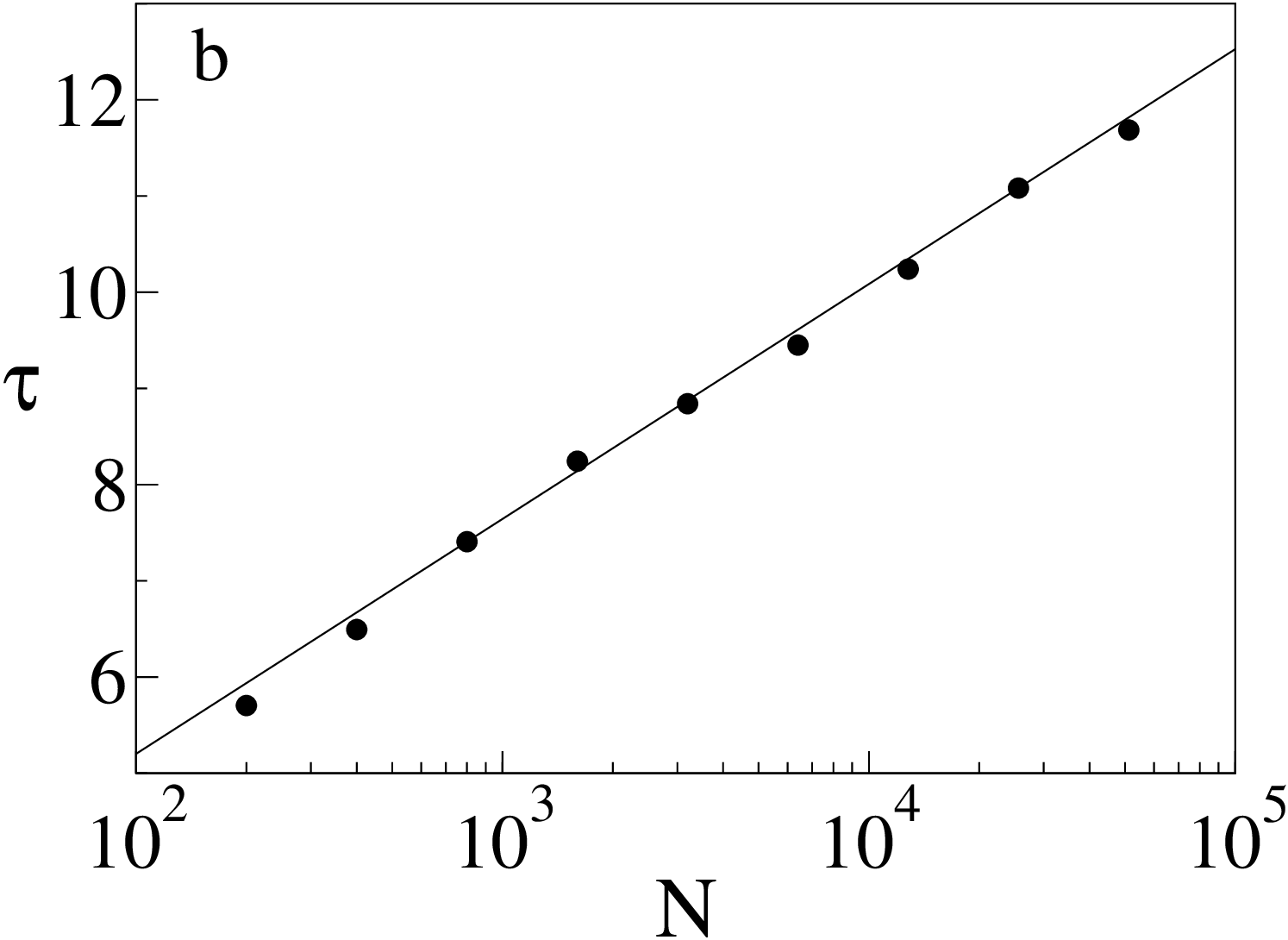}
\end{tabular}
\caption{\label{fig:metastable_stable} Relaxation time of the
$m=0$ state when this state is $(a)$ a local maximum $(K=-0.4,
\epsilon=-0.318)$ and $(b)$ a local minimum of the entropy
$(K=-0.25, \epsilon =-0.2)$.}
\end{center}
\end{figure}

We conclude this study by considering the life time $\tau (N)$ of
the $m=0$ state, when it is not the equilibrium state of the
system. In the case where $m=0$ is a metastable state,
corresponding to a local maximum of the entropy (such as in Fig.
\ref{fig:m_vs_t}a) we find that the life time satisfies $\tau \sim
e^{N \Delta s}$ where $\Delta s$ is the difference in entropy of
the $m=0$ state and that of the unstable magnetic state
corresponding to the local minimum of the entropy (see
Fig.~\ref{fig:metastable_stable}a). Such exponential dependence on
$N$ has been found in the past in Metropolis type dynamics of the
Ising model with mean field interactions~\cite{Griffiths}. Similar
behavior has been found in the $XY$ model~\cite{Torcini} and in
gravitational systems~\cite{Chavanis} when microcanonical dynamics
was applied.

We now turn to the case where the $m=0$ state is thermodynamically
{\it unstable}, where it corresponds to a local minimum of the
entropy. In systems with short range interactions, the relaxation
time of this state is finite. Here we find unexpectedly that the
life time diverges weakly with $N$, $\tau (N) \sim \log(N)$ (see
Fig.~\ref{fig:metastable_stable}b). This is to be compared with
studies of the life time of the zero magnetization state in the
$XY$ model under similar conditions which show that $\tau (N) \sim
N^{\alpha}$ with $\alpha \simeq 1.7$ \cite{Yama}.

In order to understand this behavior we consider the Langevin
equation corresponding to the dynamics of the system. The
evolution of $m$ is given by
\begin{equation}
\frac{\partial m}{\partial t}=\frac{\partial s}{\partial m} +\xi
(t) \,\,\,,\,\, <\xi(t) \xi(t')> = D \delta (t-t')
\label{Langevin}
\end{equation}
where $\xi(t)$ is the usual white noise term. The diffusion
constant $D$ scales as $D \sim 1/N$. This can be easily seen by
considering the non-interacting case in which the magnetization
evolves by pure diffusion where the diffusion constant is known to
scale in this form. Taking $s(m)=am^2$ with $a>0$, making the
$m=0$ state thermodynamically unstable, and neglecting higher
order terms, the distribution function of the magnetization,
$P(m,t)$, may be calculated. This is done by solving the
Fokker-Planck equation corresponding to (\ref{Langevin}). With the
initial condition for the probability distribution
$P(m,0)=\delta(m)$, the large time asymptotic distribution is
found to be \cite{Risken}
\begin{equation}
P(m,t) \sim \exp \left[ -\frac{ae^{-at}m^2}{D} \right] \,.
\end{equation}
This is a Gaussian distribution whose width grows with time. The
relaxation time corresponds to the width reaching a value of
$O(1)$, yielding $\tau \sim -\ln D \sim \ln N$. Similar analysis
and simulations can be carried out for the canonical ensemble
yielding the same divergence.


In summary, some general features of the dynamical behavior of
systems with long range interactions were studied using the
microcanonical local dynamics of an Ising model. Properties like
gaps in the accessible magnetization interval and breaking of
ergodicity in finite systems have been demonstrated. We also find
that the relaxation time of an unstable $m=0$ state, corresponding
to a local minimum of the entropy, is not finite but rather
diverges logarithmically with $N$. We expect these phenomena to
appear in other systems with long range interactions which are not
necessarily of mean field type. This study is thus of relevance to
a wide class of physical systems, such as dipolar systems, self
gravitating and Coulomb systems, and interacting wave-particle
systems.

We have benefited from discussions with F. Bouchet, A. Campa, T.
Dauxois, A. Giansanti, and M. R. Evans. This study was supported
by the Israel Science Foundation (ISF), the PRIN03 project {\it
Order and chaos in nonlinear extended systems} and INFN-Florence .
D.M. and S.R. thank ENS-Lyon for hospitality and financial
support.


\begin{thebibliography}{999}

\bibitem{LesHouches} T. Dauxois, S. Ruffo, E. Arimondo, M. Wilkens (Eds.),
\emph{Dynamics and Thermodynamics of Systems with Long-Range
Interactions}, Lecture Notes in Physics {\bf 602},
Springer-Verlag, New York, 2002.

\bibitem{Barre} J. Barr\'e, T. Dauxois, G. De Ninno, D. Fanelli
and S. Ruffo, Phys. Rev. E, {\bf 69}, R045501 (2004).

\bibitem{Padmanabhan} T. Padmanabhan, Phys. Rep., {\bf 188}, 285
(1990).

\bibitem{Nicholson} D. R. Nicholson, \emph{Introduction to Plasma
Physics}, Krieger Pub. Co. (1992).

\bibitem{Thirring} P. Hertel, W. Thirring, Annals of Physics, {\bf 63},
520 (1971).

\bibitem{Lynden68} V. A. Antonov, Leningrad Univ. {\bf 7}, 135 (1962); Translation in IAU
Symposium {\bf 113}, 525 (1995); D. Lynden-Bell, R. Wood, Monthly
Notices of the Royal Astronomical Society, {\bf 138}, 495 (1968).

\bibitem{Mukamel01} J. Barr{\'e}, D. Mukamel, S. Ruffo, Phys. Rev. Lett.,
\textbf{87}, 030601 (2001).

\bibitem{Yama} Y. Y. Yamaguchi, J Barr{\'e}, F. Bouchet, T. Dauxois, S. Ruffo,
Physica A, {\bf 337}, 36 (2004).

\bibitem{Celardo} F. Borgonovi, G.L. Celardo, M. Maianti and E. Pedersoli,
J. Stat. Phys., {\bf 116}, 1435 (2004).

\bibitem{Griffiths} R.B. Griffiths, C.Y. Weng and J.S. Langer, Phys. Rev., {\bf 149}, 1 (1966).

\bibitem{Nagle} J. F. Nagle, Phys. Rev. A {\bf 2}, 2124 (1970); J. C. Bonner
and J. F. Nagle, J. Appl. Phys. {\bf 42}, 1280 (1971).

\bibitem{Kardar} M. Kardar, Phys. Rev. B {\bf 28}, 244 (1983).

\bibitem{Creutz} M. Creutz, Phys. Rev. Lett. {\bf 50}, 1411 (1983)


\bibitem{Metropolis} N. Metropolis, A.W. Rosenbluth, M. N. Rosenbluth, A. H. Teller and E. Teller,
J. Chem. Phys. {\bf 21}, 1087 (1953).


\bibitem{Torcini} M. Antoni, S. Ruffo and A. Torcini, Europhys. Lett., {\bf 66}, 645 (2004).

\bibitem{Chavanis} P.H. Chavanis and M. Rieutord, Astronomy and Astrophysics, {\bf 412}, 1 (2003);
P.H. Chavanis, astrph/0404251.

\bibitem{Risken} H. Risken \emph{The Fokker-Planck Equation}, Springer-Verlag, Berlin (1996), p. 109.

\end{thebibliography}
\end{document}